\documentstyle[11pt,aas2pp4,epsfig]{article}

\def\be{\begin{equation}}
\def\ee{\end{equation}}
\def\ea{{\it et al.}\,}
\def\eg{{\it e.g.},\,}

\def\mn{{\it MNRAS}\,}

\begin{document}
\title{RXTE Spectrum of A2319}

\author{Duane Gruber\altaffilmark{1}, and Yoel Rephaeli\altaffilmark{1,2}}

\affil{$^1$Center for Astrophysics and Space Sciences, 
University  of California, San Diego,  La Jolla, CA\,92093-0424}

\affil{$^2$School of Physics and Astronomy, 
Tel Aviv University, Tel Aviv, 69978, Israel}

\begin{abstract}
The cluster of galaxies A2319 was observed in 1999 for $\sim 160$ ks by 
the PCA and HEXTE instruments aboard the RXTE satellite. No noticeable 
variability is seen in the emission measured by either instrument over 
the $\sim 8$ week observation. The quality of the data allows, for the 
first time, a meaningful search for emission whose spectral properties 
are distinct from those of the primary thermal emission with (previously 
measured) mean temperature in the range $8-10$ keV. Fitting the RXTE 
data by a single thermal component we obtain $kT = 8.6 \pm 0.1$ (all 
errors are 90\% confidence limits), a low iron abundance $Z_{F_e} \sim 
0.16 \pm 0.02$, and large positive residuals below 6 keV and between 15 to 
30 keV. The quality of the fit is {\it drastically improved} if a second 
component is added. A two-temperature model yields $kT_{1} \simeq 10.1 
\pm 0.6$, $kT_{2} \simeq 2.8 \pm 0.6$, and of $Z_{F_e} \sim 0.23 \pm 0.03$ 
which is consistent with previously measured values. An equally good fit 
is obtained by a combination of a primary thermal and a secondary 
nonthermal component, with $kT \simeq 8.9 \pm 0.6$, and power-law (photon) 
index $\alpha \simeq 2.4 \pm 0.3$. We have repeated the analysis by 
performing joint fits to both these RXTE measurements and archival ASCA 
data. At most 25\% of the RXTE secondary component could be present in the 
ASCA data. Allowing for this difference, very similar results were obtained, 
with only somewhat different values for the temperature ($kT = 9.2 \pm 0.5$) 
and power-law index ($\alpha \simeq 2.5 \pm 0.3$) in the latter model. The 
deduced value of $\alpha$ is consistent with the measured spectrum of extended 
radio emission in A2319. Identifying the power-law emission as Compton 
scattering of the radio-emitting electrons by the CMB, we obtain $B \sim 
0.1-0.3$ $\mu G$ for the volume-averaged magnetic field, and $\sim 4 \times 
10^{-14} (R/2 Mpc)^{-3}$ erg cm$^{-3}$ for the mean energy density of the 
emitting electrons in the central region (radius $R$) of A2319. 
\end{abstract}

\keywords{Galaxies: clusters: general --- galaxies: clusters: individual 
(A2319) --- galaxies: magnetic fields --- radiation mechanisms: 
non-thermal} 

\section{Introduction} 

The main goals of current work on X-ray emission from clusters of galaxies 
are refined spectral and spatial mappings to determine density and 
temperature profiles of intracluster (IC) gas, its abundance gradients, 
and to search for additional spectral components. An isothermal gas 
distribution usually provides a good overall spectral fit to the continuum 
and (usually prominent) Fe K$_{\alpha}$ line, but some temperature 
variations have already been measured in a few clusters. This is not 
surprising, as the gas distribution over typical radial regions $\simeq 2$ 
Mpc is not expected to be fully isothermal. 

In clusters with extended radio emission, there could also be an 
appreciable additional {\it nonthermal} X-ray emission, possibly from 
Compton scattering of the radio-emitting relativistic electrons by 
the Cosmic Microwave Background (CMB) radiation (Rephaeli 1979), 
or by nonthermal bremsstrahlung from a suprathermal electron population 
(Kaastra \ea 1998, Sarazin \& Kempner 2000). Measurements of clusters at 
high X energies are therefore of considerable interest: Significant 
additional insight is expected on physical conditions in the IC plasma, 
including the strength of magnetic fields, density and energy spectra of 
relativistic or suprathermal electrons and protons, and the interaction of 
these particles with the gas (Rephaeli 1979, Rephaeli \& Silk 1995). 
Search for emission at energies higher than $20$ keV can be currently made 
only with the PDS and HEXTE instruments aboard BeppoSAX and RXTE, respectively.

Initial attempts to detect nonthermal emission from a few clusters with 
the HEAO-1, CGRO, ASCA, satellites were unsuccessful (Rephaeli, Gruber 
\& Rothschild 1987, Rephaeli \& Gruber 1988, Rephaeli, Ulmer \& Gruber 
1994, Henriksen 1998). The improved sensitivity and wide spectral range of 
the RXTE and BeppoSAX satellites have recently led to major progress in the 
search for this emission. Clear evidence for the presence of a second 
component in the spectrum of the Coma cluster was seen in the analysis of 
RXTE observations (Rephaeli, Gruber, \& Blanco 1999; hereafter RGB). While 
the detection of the second component was not significant at high energies 
(where it is predicted to dominate the emission), RGB have argued that this 
component is more likely to be nonthermal, rather than a second, low 
temperature component. Indeed, this nonthermal component was directly 
detected at energies $25-80$ keV by BeppoSAX observations (Fusco-Femiano 
\ea 1999). Power-law components were also detected by BeppoSAX in two 
other clusters, A2199 (Kaastra \ea 2000), and A2256 (Fusco-Femiano \ea 2000); 
only upper limits were obtained from RXTE measurements of A2256 (Henriksen 
\ea 1999), and A754 (Valinia \ea 1999). 

Emission from nonthermal electrons may also contribute at low energies, and 
there are claims that this emission has been measured. EUV observations of 
several clusters have reportedly led to the measurement of diffuse low-energy 
($65-245$ eV) emission which is possibly nonthermal (Sarazin \& Lieu 1998, 
Bowyer \& Berghofer 1998). However, Bowyer \ea (1999) argue that this emission 
has been unequivocally detected {\it only} in the Coma cluster.

The above RXTE and BeppoSAX results provide strong motivation for extending 
the search for nonthermal components in cluster spectra. Here we present 
the results of $\sim 160$ ks RXTE observations of A2319, another cluster 
with extended radio emission region. The spectral analysis is based on the 
these measurements, as well as archival data from a $\sim 57$ ks observation 
with the ASCA satellite.   

\section{Observations and Data Reduction}
\smallskip

A2319 was observed by the Proportional Counter Array (PCA) and the High 
Energy X-ray Timing Experiment (HEXTE) on RXTE during 48 separate pointings 
totaling approximately 160\,ks between 1999 December 3, and 2000 January 20.
Spectra were accumulated by the PCA in `Standard 2' mode, which results in
a 129-channel count spectrum from 3 to 120\,keV. The HEXTE, which consists
of two independent clusters of detectors, returned data in event-by-event
mode which were subsequently accumulated into 256-channel spectra spanning
18--250 keV. To subtract the background, each HEXTE cluster was commanded
to beamswitch every 16\,s between on-source and two alternate off-source
positions 1.5$^{\circ}$ on either side. 

Standard screening criteria were applied to the data segments (Earth
elevation angle, spacecraft pointing, avoidance of the South Atlantic
Anomaly), resulting in a net exposure time of 156992\,s (as measured by the
PCA, which has negligible dead time). PCA detectors 0 and 2 were on during 
all of this time; detectors 1, 3 and 4 were enabled only occasionally, and
data from them was not used. After correcting for dead-time effects, the 
corresponding on-source live-times for the HEXTE were 61586\,s and 61818\,s 
for clusters A and B, respectively, with off-source live-times nearly equal 
to 87\% of these values.

The PCA background was estimated with the `L7/240' faint source model
provided by the instrument team. The background model appears to be
less successful at predicting the high energy ($>22$ keV) counting rate 
than the models for observing rounds 1 -- 3. The higher channels are
under-predicted by an amount which increases with energy, reaching
2\% at 120 keV.  In the following section we explicitly treat the 
uncertainty resulting from errors of background prediction and show
that it has a negligible effect on the conclusions of the analysis.
The HEXTE background was determined from the off-source
pointings, and showed no problems to a level of less than 1\%.

\section{Spectral Analysis}

We have first carried out an exhaustive analysis of the above RXTE 
observations, and then performed a joint analysis of the RXTE data with 
57 ks of public ASCA data. The 160 ks RXTE observation of A2319 provides 
a large and {\it uniform} dataset. The inclusion of ASCA measurements 
can potentially yield additional spectral insight, particularly in the 
energy range 0.8-3 keV, and some spatial information. This, however, 
comes at a price in the form of substantial additional systematics, 
such as the the energy dependence of the point spread function (\eg, 
Irwin, Bregman \& Evrard 1999). 

\subsection{RXTE Data}

For each PCA detector, the net counting rate on-source was about 14 
count/s, compared to a background rate of 13 count/s; thus A2319 
($z=0.056$) was easily detected. Since we could find no evidence for 
temporal variability in the PCA source and background rates over the 
observations, we used the time-integrated PCA and HEXTE spectra to form 
the basic data set for spectral analysis.

Response matrices were generated with standard tools, and a small allowance 
was applied for PCA systematic errors (see Wilms \ea 1999) in the
amount of 0.4\%. Additionally, PCA spectral channels below 3\,keV and 
above 22\,keV were excluded because of sensitivity to artifacts in the 
background model, and the small effective area of the PCA outside this 
range. The HEXTE data were restricted to the energy range 18--100\,keV 
for similar reasons, resulting in source and background counting rates 
of 1.2 and 128 count/s respectively (cluster~A), and 0.9 and 90 count/s 
(cluster~B). We tested for errors arising from inaccurate PCA background 
estimation in several ways, and concluded that the spectral results 
obtained here are quite robust with respect to errors of background 
determination. With well-fitting two-component models, we found that 
the best-fit model parameters changed negligibly when the PCA upper 
energy cut was extended from 22 to 60 keV. We also noted little change 
of these parameters when the background estimate was reduced by 1\% for 
better subtraction above 30 keV; this was the case with either PCA energy 
cutoff. Finally, we investigated the effect of orbital data selection by 
including more data following passage of the satellite through the South 
Atlantic Anomaly region, with its high particle fluxes. Although only 
5000 more seconds were collected, there was a dramatic increase by a 
factor of three of a barely noticeable spectral line at 16 keV. We thus 
feel that uncorrected PCA background is largely removed by the selection 
procedure, to a level which is adequately treated with the 0.4\% systematic 
error allowance. After initial testing with full energy resolution, the 
observed spectra were rebinned such that the oversampling 
was reduced to 1.5 -- 2 bins per energy resolution element, in order to 
sharpen comparison among trial spectral models in the final fitting.
Data from the individual PCA detectors was combined into a single spectrum,
and similarly a single HEXTE spectrum was prepared from the two clusters.

Most of the observed emission is clearly thermal, so we have first
determined spectral parameters of a single isothermal gas component.
The best-fit temperature (using either the MEKAL, or Raymond-Smith 
thermal emission models) in this case is $8.6 \pm 0.1$ keV, in moderate 
disagreement with the value $9.6 \pm 0.3$ keV determined from BeppoSAX 
(Molendi \ea 1999), and the value of $10.0 \pm 0.7$ keV from ASCA 
(Markevitch 1996) measurements. The observed Fe XXV K$_{\alpha}$ line 
yields an abundance $Z_{F_e} \simeq 0.16 \pm 0.02$ (in solar units), 
appreciably lower than the BeppoSAX and ASCA values, $0.25 \pm 0.03$, and 
$0.30 \pm 0.08$, respectively. No cold absorption was measurable, and given 
the 3 keV PCA threshold, none was expected. This best-fit model has an 
enormous $\chi^2$ of 216 for 31 degrees of freedom. Concave upwards 
PCA residuals indicate that an extra component is required. In particular, 
both PCA and HEXTE show strong positive residuals between 15 and 30 keV. 
When a second thermal component is added, we obtain the best-fit parameters 
(Table 1) which imply an appreciable fraction, $\sim 15\%$, for the (2 -- 
10 keV) flux in the second component.
All errors are at the 90\% confidence level, estimated by taking correlations 
between the various parameters into account (using the procedure of Lampton, 
Margon \& Bowyer 1976) for the cases of two to four `interesting' parameters. 

Equally likely is a power-law fit for the second component, shown in 
Figure 1, with statistics $\Delta\chi^2$ = 182, $\chi^2$=34, and 29 dof. 
The need for two components is overwhelmingly significant, with the
chance probability less than 10$^{-10}$. The best-fit photon index is 
$2.4 \pm 0.3$, and the temperature of the thermal component is $8.9 \pm 
0.6$ keV, not significantly different from the value for the single 
temperature fit. The deduced nonthermal 2--10 flux is $(4.0 \pm 1.0) 
\times 10^{-11}$ erg cm$^{-2}$ s$^{-1}$. With this power-law 
component, $Z_{F_e} \simeq 0.27 \pm 0.04$, in complete agreement 
with previously determined values. Figure 1 shows the fitted thermal and 
power-law components separately. The strong detection of a second 
component rests largely on the PCA spectrum from 3 to 6 keV. But the data 
{\it above 18 keV}, for which the fitted power-law flux is about equal 
that of the isothermal component, independently indicate a power-law 
second component. By fitting only HEXTE data above 18 keV with the 
isothermal parameters fixed at their values determined over the entire 
energy range, the power-law component has joint 90\% confidence intervals 
of [0.6, 1.2] in units of $10^{-11}$ erg cm$^{-2}$ s$^{-1}$ for the 18-40 
keV flux, and [1.8, 5.0] for the index. Both values are consistent with the 
globally-determined estimates. If the index is held fixed at 2.4, the 
power-law flux above 20 keV is significant at the $8 \sigma$ level. 

\begin{figure}[h]
\centerline{\psfig{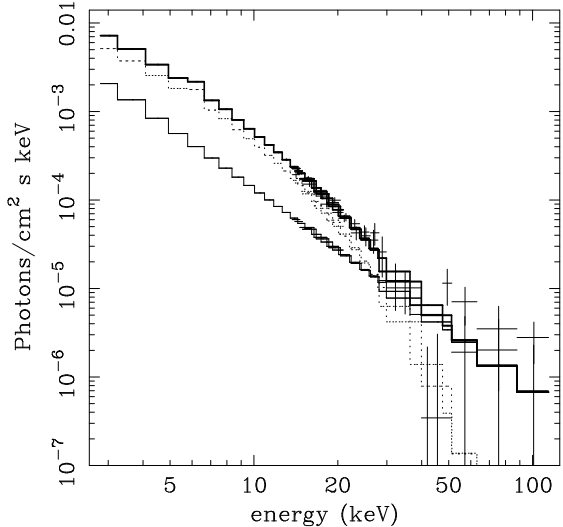}}
\figcaption{Unfolded RXTE data and combined Raymond-Smith plus power-law 
models with $8.9 \pm 0.6$ keV and index of $2.4 \pm 0.3$, shown individually, 
as well as their sum.}
\end{figure}

\subsection{Joint RXTE and ASCA Datasets}

ASCA data provide a comfortable spectral overlap with RXTE data between 3 
and 8 keV, so if the second component is not time variable, we can attempt 
to confirm it through joint fitting of both datasets. ASCA spectra were 
extracted using standard tools from the cleaned archival GIS event files. 
Extraction radii between 15 and 20 arcminutes were found to include 
essentially the whole cluster emission. We chose 0.8 -- 8.0 keV as the most 
reliable band for ASCA spectra, and grouped the hundreds of energy channels 
in this range into 21 spectral bins, each of width $ \delta E/E$ = 0.1. 
Data from the two GIS detectors were also combined. As with the RXTE PCA, we 
added systematic noise of about 1\% per final spectral bin.

Fitting a single Raymond-Smith spectral form to the ASCA data alone, we 
obtained a temperature of 12.5 keV, considerably higher than the 10 keV 
reported by Markevitch (1996). By narrowing the extraction radius to the 
6 arcminutes that is used for point sources, we obtained a kT of 9 keV, 
much more consistent with temperatures obtained by Markevitch (1996), 
BeppoSAX (Molendi \ea 1999), and here with RXTE. We then checked the 
spectral inter-calibration of ASCA and RXTE, as well as the standard 
analysis tools, on the Crab. With both ASCA and RXTE we obtained a 
spectral index of 2.08. Thus we were very confident of the on-axis 
effective area calibration of ASCA, but felt that the effective area at 
5 -- 15 arcminutes off-axis may be underestimated by about 10\% at 
higher energies. Because of the exact agreement on the Crab, we employed 
the ASCA data extracted with 5 arcminute radius for the joint analysis 
with RXTE. With this radius a fraction 0.59 of the cluster flux is 
collected, so a normalization constant of 0.59 for the ASCA spectrum was 
initially used in the analysis. When allowed to float in the joint analysis, 
values near 0.6 are in fact obtained.

\begin{figure}[h]
\centerline{\psfig{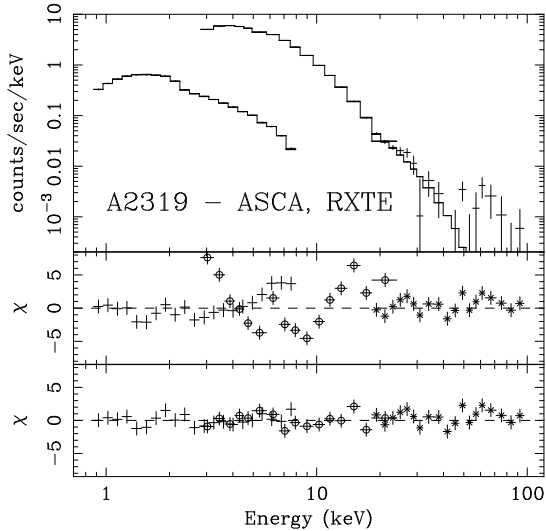}}
\figcaption{Joint fit to ASCA and RXTE data with a common Raymond-Smith
thermal model and an RXTE-only power-law component. The displayed
data points have been combined from adjacent energy channels
of the two ASCA GIS detectors, two RXTE PCA detectors (circles)
and the two RXTE HEXTE clusters (crosses). The upper panel
shows the data and best fit, the middle panel shows large
residuals to the best isothermal fit, and the bottom panel
shows acceptable residuals after a power law component is added
for the RXTE data.}
\end{figure}

In the joint fits the main component was modeled as a Raymond-Smith 
plasma emission; a Mekal model produced nearly identical results in 
all fits. A fit with a single temperature had $\chi^2$ of 313 for 
50 degrees of freedom. The RXTE PCA data (see Figure 2, middle panel) 
were mainly responsible for this very poor fit. Two-temperature models
produced negligible improvement in $\chi^2$. Likewise, a thermal plus 
power-law model produced little improvement in $\chi^2$. Thus, the ASCA data 
do not confirm the power-law or low-temperature thermal component so strongly 
indicated by the RXTE data alone. Nevertheless, when the condition of equal 
RXTE and ASCA intensity for the second component is relaxed, reduced $\chi^2$ 
drops from the value 6 to near unity. In these best-fit models the flux of the 
secondary component is not significant in the ASCA data. Completely 
acceptable $\chi^2$ values were achieved by fitting an additional power-law or 
thermal model to the RXTE data. With the inclusion of the ASCA data the 
best-fit parameters differ little from the RXTE-only fits, as shown in 
Table 1. The temperature and abundance of the primary component are unchanged 
at 10.1 keV, and 0.23, respectively. Fitting the second component with a 
power-law model changes the results mildly, with kT raised from 8.9 to 9.2 
keV, $Z_{F_e} \simeq 0.29$ instead of 0.27, and a power-law index of 2.5 
rather than 2.4. Errors are only slightly reduced from their respective 
RXTE-only values. 90\% confidence upper limits of 25\% were obtained to 
the fraction of the RXTE secondary component -- power-law or thermal -- 
that could be present in the ASCA data.

\begin{table*}
\caption{Results of the spectral analysis}
\bigskip
\begin{tabular}{|ll|ccc|}
\hline
Dataset & Parameter & Isothermal  & Two-Temperature & Isothermal + Power-law \\
\hline
RXTE & $kT_1$    & 8.6$\pm$0.1 & 10.1$\pm$0.6   & 8.9$\pm$0.6  \\

     & Normalization$^a$  & 0.171$\pm$0.002 & 0.138$\pm$0.025 & 0.129$\pm$
0.010 \\

     & $kT_2$ (keV)       &         & 2.8$\pm$0.6    &               \\

     & Normalization$^a$  &           & 0.059$\pm$0.010 &     \\

  & $f$(2-10\,keV) (erg $cm^{-2}\,s^{-1}$) & & (1.10$\pm 0.27)\times 10^{-4}$\\

     & Photon index      &             &                & 2.4$\pm$0.3    \\

     & Fe abundance$^b$     & 0.16$\pm$0.02 & 0.23$\pm$0.03 & 0.27$\pm$0.041 \\

     & $\chi^2/dof$      & 216/31     & 34/29   & 34/29         \\
\hline

RXTE+ASCA & $kT_1$  & 8.6$\pm$0.1 & 10.1$\pm$0.7   & 9.2$\pm$0.5  \\

    & Normalization$^a$ & 0.172$\pm$0.002 & 0.140$\pm$0.014 & 0.127$\pm$0.012\\

    & $kT_2$ (keV)       &         & 2.4$\pm$1.0    &               \\

    & Normalization$^a$  &           & 0.061$\pm$0.010 &     \\

 & $f$(2-10\,keV) (erg $cm^{-2}\,s^{-1}$) & & & (4.0$\pm 1.0)\times 10^{-11}$ \\

    & Photon index      &             &                & 2.5$\pm$0.3  \\

    & Fe abundance$^b$     & 0.16$\pm$0.02  & 0.23$\pm$0.03 & 0.29$\pm$0.04\\

    & $\chi^2/dof$      & 313/50     & 49/48   & 50/48   \\
\hline
\end{tabular}
\tablenotetext{}{Notes: }
\tablenotetext{}{All quoted errors are at the 90\% confidence level. }

\tablenotetext{a}{e.m. = Raymond-Smith emission measure in units of
10$^{-14} \int N_eN_H dV$ / 4$\pi D^2$, where $D$ is the luminosity distance
and $N_e$, $N_H$ are the total number of electrons and protons, respectively.}

\tablenotetext{b}{Abundance is expressed relative to solar values.}
\end{table*}

\section{Discussion}

The above spectral analysis of the RXTE measurements of A2319 shows clearly 
that the emission at ($\epsilon >3$ keV) cannot be adequately described in 
terms of a single isothermal gas component. A second component is not 
required to fit the archival ASCA data. Above all this reflects the fact 
that the detection of a second component in cluster spectra is still a 
challenging task. Because of the much longer exposure of the RXTE 
observation, and the much wider spectral range of PCA and HEXTE, and given 
the substantial systematic uncertainties associated with the analysis of 
joint datasets from two satellites, we consider the spectral results of 
the RXTE analysis to be more robust. However, the much larger RXTE field 
of view, and its lack of spatial resolution, constitute sufficient ground 
for treating our main result as tentative, requiring independent 
confirmation by an X-ray satellite with the requisite wide spectral 
response and spatial resolution of at least $\sim 10'$. In the rest of 
this section we consider some of the implications of the results from the 
RXTE data analysis.

The measured RXTE spectrum can be well fit either by a two-temperature 
plasma, or by a sum of isothermal and power-law emissions. Essentially the 
same results were obtained in our analysis of the $\sim 90$ ks RXTE 
observation of the Coma cluster (RGB). The secondary emission component in 
Coma was found to contribute only $\sim 8\%$ of the total 2-10 keV emission; 
in A2319, this contribution is significantly higher -- $\sim 15\%$ for 
a thermal, and $\sim 22\%$ for a power-law, secondary components.

The characterization of the thermal state of IC gas over a large (several 
Mpc) region by a single isothermal distribution is clearly simplistic, so 
it is not surprising that detailed, high quality observations with previous 
(\eg Markevitch 1996, Honda \ea 1996) and current X-ray satellites, 
indicate the need for more realistic spectral modeling. What is currently 
feasible is the measurement of appreciable emission with spectral 
properties sufficiently different from those of the primary thermal 
component with which the emission has been hitherto identified. Thus, 
non-isothermality of IC gas can be characterized in first order by a second 
thermal component with a temperature different from the main component. 
Alternatively, a secondary component could have a power-law form, 
associated with emission from a nonthermal distribution of energetic 
electrons in either powerful radio sources (and other active galaxies), 
or in extended regions of the IC space. The best-fit analysis presented 
above confirms the likelihood of both these two spectral possibilities, 
which - from a formal statistical point of view - have been found to be 
both likely. Let us first consider some of the physical implications 
from the possible presence of each of these spectral components. It 
may be considered natural to expect that a second emission component 
is also thermal, and that this is just a realistically testable 
manifestation of what is actually a continuous temperature distribution 
in the IC gas. This by itself is almost self-evident; what may be unclear, 
however, is whether $\sim 15\%$ of the deduced 2-10 keV flux from 
$kT_{2} \sim 2 --3$ keV gas is physically likely, and whether such 
emission is consistent with previous measurements which could have probed it. 

There is at best only some indication for a temperature gradient in the 
central region of A2319. An average temperature, $kT \simeq 10.0 \pm 0.7$ 
keV, was determined from ASCA measurements (Markevitch 1996) of the central 
$\sim 15'$ radial region. In most of this region the temperature was found 
to be closer to 11 keV, whereas a lower ($kT \simeq 8.4 \pm 1.2$ keV) value 
was measured within a subcluster area. BeppoSAX/MECS measurements lead to a 
similar value of $kT \simeq 9.6 \pm 0.3$ keV for the mean temperature in 
roughly the same region; no evidence is seen in these data for a 
significant decrease in the temperature. However, the effective radius of 
the RXTE FOV is about twice the size of the region probed by the ASCA and 
MECS measurements, so a large-scale temperature gradient is perhaps even 
more likely to be detected by the RXTE. It is of interest, therefore, to 
determine the temperature profile that is consistent with our results, and 
we did so in the context of polytropic gas models, adopting the approach 
developed by RGB in their similar analysis of RXTE measurements of the 
Coma cluster. As in the case of the latter cluster, no acceptable combination 
of the relevant parameter values (polytropic index and the density profile 
$\beta$ parameter) is found to yield values of the temperatures and fluxes
which are consistent with the results of the observations. Invariably, very 
high central temperatures are required in order to have substantial low 
temperature emission in the outer region of the cluster. This is in direct 
conflict with the results obtained from the high spatial resolution ROSAT, 
ASCA, and BeppoSAX/MECS measurements cited above. The inconsistency of 
polytropic temperature profiles with the ROSAT results was previously 
noted also by Trevese, Cirimele, \& Simone (2000).

Can the two spectral components be explained as emission from lower 
temperature gas clumps surrounded by hotter diffuse gas? The possibility 
that IC gas may be appreciably clumped, to the extent that its emission 
characteristics differ in a measurable way from those of uniformly 
distributed gas, was previously investigated in the context of an isobaric 
equilibrium model for the two components. Various considerations lead to 
the conclusion (e.g., Rephaeli \& Wandel 1984, Holzapfel \ea 1997) that 
only a very small fraction of the gas can survive in small clumps over 
relevant cluster timescales ($> 10^{9}$ yr). This fraction is too small to 
account for nearly 15\% of the total flux from the clouds.
We therefore conclude that even though IC gas may not be fully isothermal, 
the deduced level of additional low-temperature emission required to fit 
the RXTE data is considerably higher than can be readily explained by 
either polytropic, or clumpy gas distributions. 

The possibility that the additional spectral component is nonthermal was 
found to be statistically about equally likely, as quantified in the 
previous section. Consider now the physical implications from the 
substantial power-law component that we have deduced. We first note that 
there is one known AGN, RXJ1923.1+4341, at a projected distance of 34' from 
the center of A2319, which was observed in surveys by both the 
{\it Einstein} and ROSAT satellites. The measured (IPC and PSPC) fluxes 
differ by at most a factor of 2, but are at least a factor $\sim$60 lower 
than the flux deduced here in the second component of A2319. It is therefore 
unlikely that the emission is from bright sources in the RXTE field of view. 
The fact that no temporal variability is seen in our data provides additional 
evidence for no appreciable AGN contribution, but does not, of course, rule 
it out.

An eminently reasonable possibility is that the nonthermal emission results 
from Compton scattering of the radio-producing electrons by the CMB. 
Diffuse radio emission from A2319 was measured at 26 and 610 MHz (Erickson, 
Mathews, \& Viner 1978; Harris \& Miley 1978) from a central $20^{\prime}$ 
region. Feretti, Giovanini,\& B\"ohringer (1997) have recently reported the 
results of WSRT and VLA observations at 323, 327, 1420, and 1465 MHz. The
emission is found to be more extended than previously determined (and also 
more powerful than that in the Coma cluster). Properties of the emission 
vary considerably across the radio emitting region, particularly the 
spectral index, for which the range $\sim 0.9 - 2.2$ was deduced 
(Feretti \ea 1997; see also Molendi \ea 1999). Diffuse IC emission 
is expected to have a steeper spectrum than in galaxies (whose indices 
are typically in the rough range $\sim 0.7-0.8$) due to continued 
aging resulting from Compton-synchrotron losses over (long) propagation 
times to the IC space. This index is thus predicted to be steeper by $\sim 
1/2$, as has been found to be the case in the Coma cluster. (A lower 
measured value might indicate incomplete subtraction of galactic emission.) 
The measured radio index is thus quite uncertain, but can roughly be 
taken as $1.5 \pm 0.6$. Perhaps a more useful characterization of the 
spectrum is its flux level of 1.45 Jy at 408 MHz (Feretti \ea 1997).

Radio measurements imply the presence of electrons with energies of at 
least few GeV. Compton scattering of the electrons by the CMB boosts 
photon energies to $\sim 1-100$ keV, and probably even higher. The 
spectral (photon) flux depends strongly on the value of the mean, 
volume-averaged magnetic field, $B$. If we identify this emission as 
due to Compton scattering by the same 
electrons that give rise to the observed radio emission, then the implied 
radio index would be $\simeq 1.5 \pm 0.3$. With this value of the index, 
the deduced power-law flux, and the measured flux at 408 MHz, we can 
compute $B$, using the usual Compton-synchrotron formulae (\eg Rephaeli 
1979). Doing so we estimate $B$ to be in the range $\sim 0.1 - 0.4$ 
$\mu$G, when account is taken of the combined uncertainty intervals of 
the coefficient and index in the power-law X-ray flux. While the mean 
value of the magnetic field is independent of the source size and 
distance (when it is assumed that the radio and HEX sources have the 
same size), the relativistic electron energy density, $\rho_{e}$, does 
depend on these quantities, and the range of electron energies. With 
$B = 0.1 - 0.4$ $\mu$G, and integrating the electron energy distribution 
over energies in the observed radio and X-ray bands, we obtain $\rho_{e} 
\sim (2 - 5) \times 10^{-14} (R/(2 \, Mpc))^{-3}$ erg\,cm$^{-3}$ for the 
mean electron energy density in a 2 Mpc spherical region at a (luminosity) 
distance of $341$ Mpc (with 

We emphasize that due to the lack of spatial information our estimates of 
the the mean value of the magnetic field, and especially the electron 
energy density, are just rough averages. In particular, we caution against 
drawing definite conclusions based on the disparity between the generally 
lower field values determined from the combination of X-ray and radio 
observations (Rephaeli, Gruber \& Blanco 1999, Fusco-Femiano \ea 1999), 
and field values determined from Faraday rotation measurements (see, \eg 
Kim, Tribble \& Kronberg 1991, Clarke, Kronberg, \& B\"ohringer 2001). 
For a more extensive discussion of this issue, see Goldshmidt \& Rephaeli 
(1993), and Newman, Newman \& Rephaeli (2001).

As noted in the Introduction, nonthermal emission in the observed energy 
range can possibly be produced also by nonthermal bremsstrahlung of 
suprathermal electrons (Kaastra \ea (1998, Sarazin \& Kempner 2000). 
However, this interpretation requires the identification of a second 
energetic electron population which is distinct from the relativistic 
population that produces the observed radio emission. The determination 
of the exact properties of the energetic electron energy distribution 
would necessitate more accurate spectral and spatial measurements.

The flux in the nonthermal component reported here is comparable to the 
limit obtained from the BeppoSAX PDS measurements (Molendi \ea 1999); they 
report an upper bound (at 90\%) of $\sim 2.3 \times 10^{-11}$ 
erg cm$^{-3}$ s$^{-1}$ on the flux in the PDS band (13-200 keV), as 
compared with our measured value of $\sim (2.5 \pm 0.36) \times 10^{-11}$ 
erg cm$^{-3}$ s$^{-1}$ (both at 90\% confidence). The nonthermal X-ray 
luminosity (taking a luminosity distance of $\simeq 341$ Mpc) in the same 
band is quite substantial, $\sim (3.5 \pm 0.5) \times 10^{44}$ 
erg s$^{-1}$, when compared with the bolometric thermal luminosity, 
$\sim 2.5 \times 10^{45}$ erg s$^{-1}$.

In conclusion, deep exposure of A2319 with the RXTE satellite provides 
strong evidence for a second component in the spectrum of A2319. While 
from merely a statistical point of view there is no preference in the 
data for a power-law as compared to thermal secondary emission, the 
very low temperature deduced for the second component is both observationally 
and theoretically problematic, whereas nonthermal emission is expected 
based on radio measurements. It will be of great interest to observe A2319 
with the IBIS instrument aboard the upcoming INTEGRAL satellite. The IBIS 
broad spectral band and imaging capabilities will provide essential 
information on the spatial distribution of the high-energy emission from 
the cluster. This will likely lead to an unequivocal identification of 
the nature of the second spectral component.

\acknowledgments
The scope of the work presented in this paper was expanded by performing 
a joint RXTE and ASCA data analysis; we thank the referee for urging us 
to do so.

\parskip=0.02in
\def\ref{\par\noindent\hangindent 20pt}
\noindent
{\bf References}
\smallskip

\ref{Bowyer, S., \& Berghofer, T.W. 1998, ApJ, 506, 502}

\ref{Bowyer, S., \ea 1999, ApJ, 526, 592}

\ref{Clarke, T.E., Kronberg, P.P., \& B\"ohringer, H. 2001, ApJ, 547, 
L111}  

\ref{Erickson, W.C., Matthews, T.A., \& Viner, M.R. 1978, ApJ, 222, 761}

\ref{Feretti, L., \ea 1997, New Astron., 2, 501}

\ref{Fusco-Femiano, R. \ea 1999, ApJL, 513, L21}

\ref{Fusco-Femiano, R. \ea 2000, ApJL, 534, L7}

\ref{Giovannini, G., \ea 1993, ApJ, 406, 399}

\ref{Goldshmidt, O., \& Rephaeli, Y., 1993, ApJ, 411, 518}

\ref{Harris, D.E., \& Miley, G.K. 1978, A\&A Supp., 34, 117}

\ref{Henriksen, M. 1998, ApJ, 50, 389}

\ref{Henriksen, M., \ea 1999, ApJ, 511, 666}
 
\ref{Holzapfel, W.L., \ea 1997, ApJ, 480, 449}

\ref{Honda, H., \ea 1996, ApJ, 473, L71}

\ref{Irwin, J.A., Bregman, J.N., \& Evrard, A.E. 1999, ApJ, 519, 518}

\ref{Kaastra, J.S. \ea 1998, Nuc. Phys. B, 69, 567}

\ref{Kaastra, J.S. \ea 2000, ApJL, 519, L119}
 
\ref{Kim, K.T., \ea 1990, ApJ, 355, 29}

\ref{Lampton, M., Margon, B., \& Bowyer, S. 1976, ApJ, 208, 177}

\ref{Markevitch, M. 1996, ApJ, 465, L1}

\ref{Molendi, S., \ea 1999, ApJ, 525, L73}

\ref{{Newman, W.I, Newman, A.L., \& Rephaeli, Y. 2001, preprint}

\ref{Wilms, J., \ea, 1999, ApJ, 522, 460}

\ref{Rephaeli, Y. 1979, ApJ, 227, 364}

\ref{Rephaeli, Y., Gruber, D.E., \& Rothschild, R.E. 1987, ApJ, 320, 139}

\ref{Rephaeli, Y., \& Gruber, D.E. 1988, ApJ, 333, 133}

\ref{Rephaeli, Y., Gruber, D.E., \& Blanco, P.R. 1999, ApJ, 511, L21}

\ref{Rephaeli, Y., \& Silk, J. 1995, ApJ, 442, 91}

\ref{Rephaeli, Y., Ulmer, M., \& Gruber, D.E. 1994, ApJ, 429, 554}

\ref{Rephaeli, Y., \& Wandel, A. 1984, \mn, 215, 453}

\ref{Sarazin, C.L., \& Kempner, J.C. 2000, ApJ, 533, 73}

\ref{Sarazin, C.L., \& Lieu, R. 1998, ApJ, 494, L177}

\ref{Valinia, A. \ea 1999, ApJ, 515, 42}

\end{document}